\documentclass{article}
\usepackage[utf8]{inputenc}
\usepackage{graphicx}
\usepackage{epsfig}
\usepackage{amsmath}
\usepackage{amssymb}
\usepackage{braket}
\usepackage{indentfirst}
\usepackage{appendix}

\title{Feynman diagram approach to dynamical Casimir effect in optimechanical cavity}
\date{}
\author{Yu-Song Cao\footnote{caoyusong15@mails.ucas.ac.cn}$~^{1,2}$,YanXia Liu$~^{2}$}

\begin{document}
\maketitle

\noindent $~^{1}$\small{School of Physical Sciences, University of Chinese Academy of Sciences, Beijing, 100049, China}

\noindent $~^{2}$\small{Beijing National Laboratory for Condensed Matter Physics, Institute of Physics, Chinese Academy of Sciences, Beijing 100190, China}

\begin{abstract}
    In this paper we study an optomechnical system enclosed by an optical cavity with one mirror attached to a spring as a closed quantum system. We provide a different angle of studying the phenomenons related to the dynamical Casimir effect via Feynman diagram technique. Dressing effects of phonon, photon and coupling strength are discussed. The energy shift of ground state is obtained. The dynamical Casimir effect is modeled by the scattering processes converting phonons to photons and the corresponding scattering amplitudes are computed. The force-force correlation function of the radiation pressure is derived, whose non-Gaussian probability distribution is revealed.
\end{abstract}

\section{Introduction}

        The magic lies in the center of quantum mechanics is the uncertainty principle, as a result of which, the quantum systems exhibit non-vanishing zero-point energy. When the degrees of freedom of the system goes infinite, we encounter the first intrinsic divergence in quantum field theory: the absolute value of vacuum energy \cite{Itykson}. This crisis is solved by introducing the normal ordering technique, which indicates the absolute value of vacuum energy has no physical meaning because it is unobservable. However, it is not the case for the real world. In gravitational systems, the Einstein tensor is coupled to the energy momentum tensor thus the vacuum energy gives rise to the accelerating expansion of the universe \cite{cosmo}. The vacuum energy shows its presence also in the flat timespace, it is discovered that if the quantum vacuum is disturbed non-adiabatically, particles can emerge from the vacuum, which gives rise to the astonishing phenomenons such as dynamical Casimir effect \cite{RevofMod,vacuum,moore,haro}, Hawking radiation \cite{HR1,HR2,cosmological} and spontaneous radiation \cite{Itykson}. So far, the dynamical Casimir effect has been verified in many experimental systems \cite{63,A100}. It is notable that the dynamical Casimir effect serves as an analog of studying the Unruh effect \cite{UE} and Hawking radiation \cite{RevofMod,D36,D37,D77}.

        Many theoretical treatments of the dynamical Casimir effect are developed semiclassically, where the motion of the boundary is given classically prior and the field dynamics is described quantized field equation \cite{Neto96,scattering,82,Neto94,Unruh14,Unruhmassive}. Although the semiclassical theory proved its success, however, people still want a fully quantized description of dynamical Casimir effect. The first reason is philosophical consistence while the second reason is more realistic. In systems requiring the high accuracy of the measurement of the mirror's position, the quantum fluctuation must be taken into consideration \cite{LIGO}. Moreover, the semiclassical theory has an Achillies' heel of failing to give an integrated description of dynamical properties of the boundary. In semiclassical theory, the backaction force quantum field exerts on the mirror is given in an average level \cite{scattering,82,Neto94,Unruh14,Unruhmassive,1998}.

        The fully quantum description of the dynamical Casimir effect is made by Law \cite{Law95}. After which various new effects as a result of the quantum nature of the mirror have been discovered \cite{EPL,Italian,A58,D91,D96,L111,Argentina,PRX}. The backaction effect, is of course studied. In \cite{EPL,Italian}, the dynamical Casimir effect is modeled by the evolution of the mean value of the field operator and excitation number of the harmonic oscillator and the dispassion to the enviorment is considered crucial. In this paper, we study the behavior of Hamiltonian in \cite{Law95} with Feynman diagram approach. The Feynman rules with single mode approximation is obtained, which will show its power in studying the dynamical effects of the system. With Feynman diagrams we study the dressing effects of the photon, phonon and the coupling strength. We give a picturise description on how the virtual particles presents in physical photon as well as physical phonon. We also verify the dressing effect of vacuum, where the non-vanishing mean value of photon number in ground state will modify the ground state energy \cite{L111}. Beyond the single mode approximation, we briefly discuss the effective interaction between photon modes induced by the fluctuation of phonon vacuum. The dynamical Casimir effect is modeled by the scattering processes converting phonon to photons, and the corresponding transition amplitudes are obtained. We also derive the correlation function of the backaction force, which is non-Gaussian distributed in our model. And we also point out that the correlation function of backaction force is very small, meaning the Langivan-type equation serves as a good approximmation describing the mirror's motion. Throughout this paper the systems is considered in weak coupling regime and the dispassion is disregarded.

\begin{figure}
\begin{center}
\includegraphics[width=.42\textwidth]{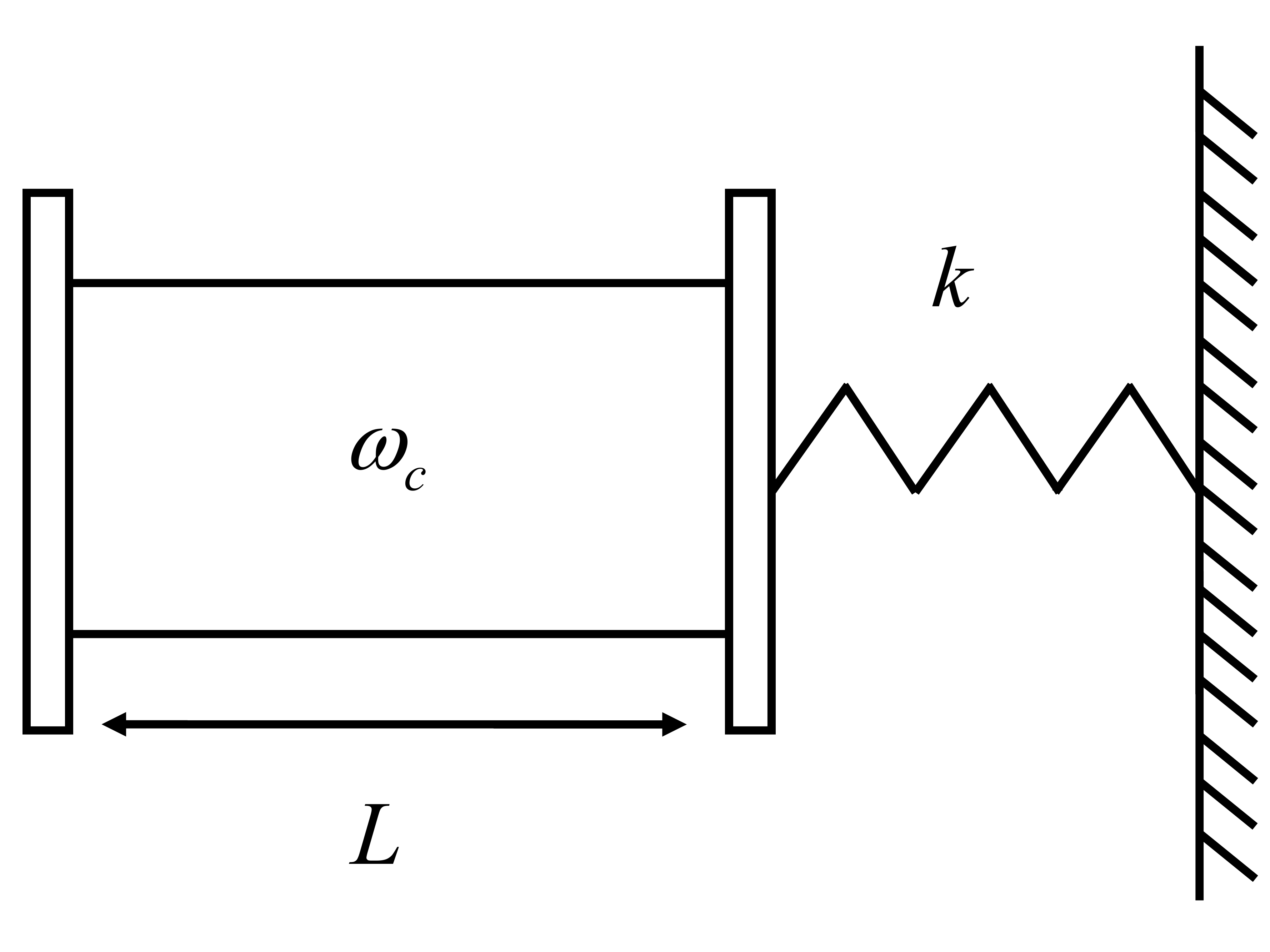}
\caption{The sketch of our system: An optical cavity with one fixed mirror and another attached to a spring. The cavity field acts radiation pressure on the mirror and the motion of the mirror squeezes the cavity field.}
\label{fig:dem}
\end{center}
\end{figure}

        This paper is organized as follows. In Sec.2, we give a brief introduction of the model Hamiltonian and present the Feynman rules. In Sec.3, we study the dressing effects of phonon, photons and the coupling strength. We also demonstrate the existence of the virtual particles in vacuum state, which shifts the ground state energy of the system. The vacuum induced photon-photon interaction is also briefly discussed. In Sec.4, we compute the transition amplitudes of the processes modeling dynamical Casimir effect and derive the correlation function of the backaction force, whose non-Gaussian possibility distribution is revealed. In Sec.5, we give a conclusion of our results and give some discussion.

        Throughout this paper the units are chosen as $\hbar=c=1$.

\section{Model Hamiltonian and Feynamn rules}
        The schematic of the system is demonstrated in Fig.(\ref{fig:dem}). The mass of the movable mirror is $m$ and the strength of the spring is $k$. $L$ is the length of the cavity when the spring is at relaxation. Under the action of the spring, the mirror behaves like a simple harmonic oscillator and the radiation pressure on the mirror can be regarded as driving force. In the meantime, the motion of the mirror will also squeeze the cavity field non-adiabatically. In our discussion, the radiation pressure is small compared to the elastic force from the spring, thus the mirror's sphere of activity is much smaller compared to $L$.

        The Hamiltonian of this system is given as $H=H_{0}+V$, with the free term
        \begin{equation}\label{eq:H0}
        H_{0}=\omega_{c}a^{\dagger}a+\omega_{m}b^{\dagger}b,
        \end{equation}
        where $a$ is the cavity field operator whose frequency is $\omega_{c}=n{\pi}/L$, with $n$ for the arbitrarily chosen mode number, and $b$ is the annihilation operator of the harmonic oscillator of form $b=\sqrt{\frac{m\omega_{m}}{2}}(x+\frac{ip}{m\omega_{m}})$, with $x$ and $p$ being the position and momentum operators of the harmonic oscillator, thus the frequency of harmonic oscillator is given by $\omega_{m}=\sqrt{\frac{k}{m}}$. Note that here we use the single mode simplification which will be sufficient for making our point. Meanwhile, there are two polarizations of the electromagnetic field inside the cavity, which gives only a redundant freedom in the effects we will discuss later \cite{Argentina}. So we will consider only one polarization in this case. Throughout this paper, the exciton number of the harmonic oscillator is referred as phonon number \cite{RevofMod}.

        The radiation pressure on the mirror is given by \cite{EPL,Italian}
        \begin{equation}\label{eq:force}
        F=\frac{\omega_{c}}{2L}(a+a^{\dagger})^{2},
        \end{equation}
        from which we obtain the interaction term as
        \begin{equation}\label{eq:V}
        \begin{split}
        V&=Fx\\
        &=ga^{\dagger}a(b+b^{\dagger})+\frac{g}{2}[a^{2}+(a^{\dagger})^{2}](b+b^{\dagger}),
        \end{split}
        \end{equation}
        where the normal ordering technique is applied \cite{L111} and
        \begin{equation}
        g=-\sqrt{\frac{1}{2m\omega_{m}}}\frac{\omega_{c}}{L}.
        \end{equation}
        The interaction term is split into two parts: $V=V_{om}+V_{DCE}$, the first part
        \begin{equation}
        V_{om}=ga^{\dagger}a(b+b^{\dagger})
        \end{equation}
        is called the standard optomechanical coupling term which conserves the number of photon \cite{PRX} and the second term
        \begin{equation}
        V_{DCE}=\frac{g}{2}[a^{2}+(a^{\dagger})^{2}](b+b^{\dagger})
        \end{equation}
        is called dynamical Casimir effect term being responsible for the dynamical Casmir effect. In this paper, the cavity length $L$ is a macroscopic quantity and mode number $n$ is taken not very large. Thus we obtain the weak coupling condition $\omega_{c},\omega_{m}{\gg}g$. In experimentation, the optomechanical coupling term $V_{om}$ plays the dominant role and the effects from dynamical Casimir effect term is hard to observe when the coupling $g$ is too small \cite{PRX}.

\begin{figure}
\begin{center}
\includegraphics[width=.55\textwidth]{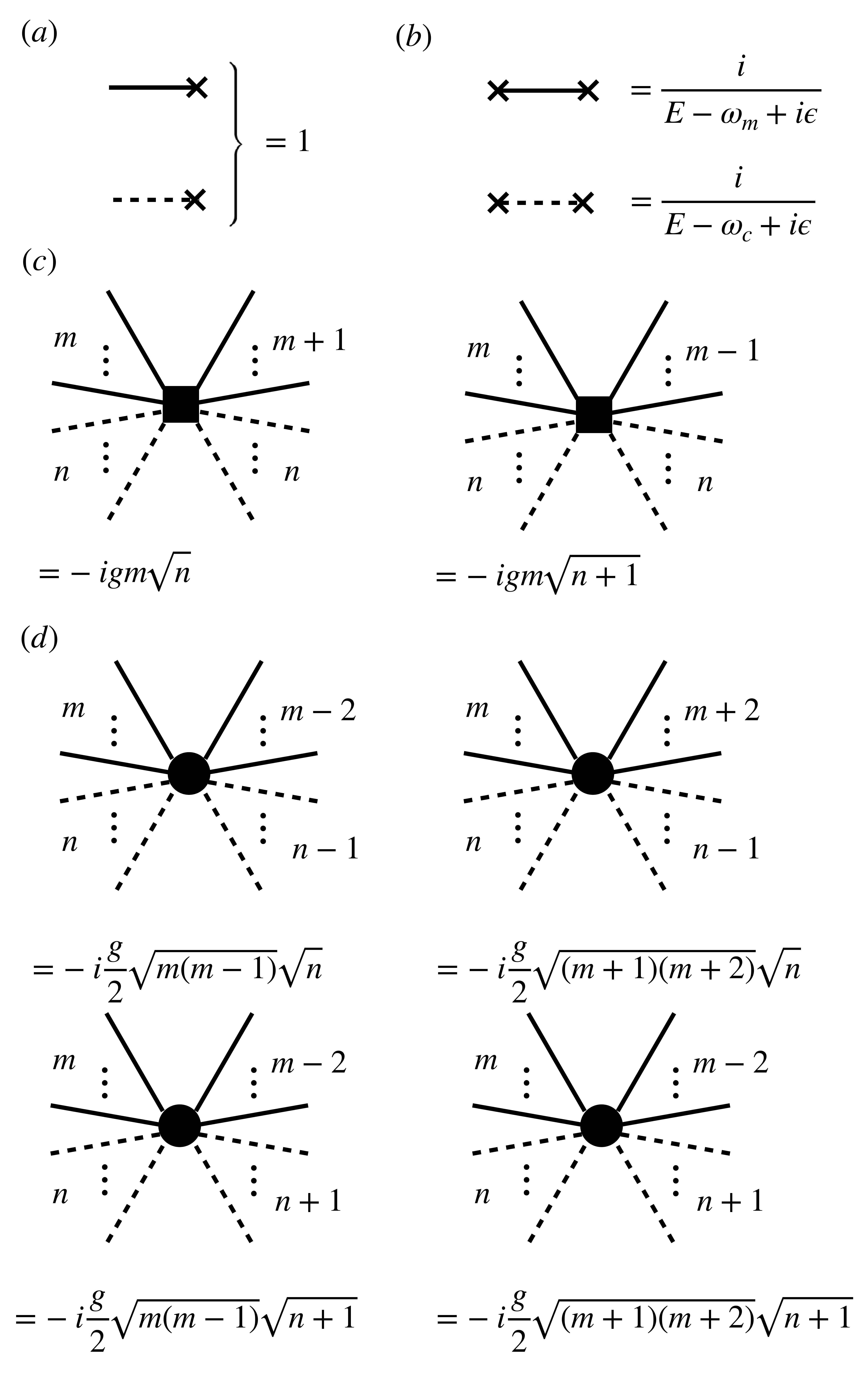}
\caption{Feynman rules in momentum space. (a) External lines. (b) Propagators. (c) Optomechanical interaction vertices, the numbers denote the number of the lines attached to the vertex. (d) Dynamical Casimir effect interaction vertices, the numbers denote the number of the lines attached to the vertex. Note that the cross mark in (a) and (b) can be any type of vertices. And the boson symmetry factor is absorbed in the interaction strength in (c) and (d). In (a) and (b), the legs are drawn short to save space thus we name them "Corgidiagrams".}
\label{fig:FR}
\end{center}
\end{figure}

        Like the case of Rabi model \cite{Stephano}, this optomechanical system can also be viewed as an anology to quantum electrodynamics. Thus we are encouraged to apply the Feynman diagram technique. Since we are working in non-relativistic regime, the field propagators goes only from the backward towards the future, making us unable to use Wick's theorem. After some algebra, the Feynman rules is obtained and is shown in Fig.(\ref{fig:FR}), whose detailed derivation is given in Appendix A. Note that the Bose symmetry factor is absorbed into the coupling strength in Fig.(\ref{fig:FR}). From Fig.(\ref{fig:FR}) we can see the two kinds of interaction in Eq.\eqref{eq:V} gives two types of the vertices, the optomechanical ones are marked by squares while the dynamical Casimir effect ones marked by circles. Note that the different kinds of interaction vertices give rise to different value of coupling strength, as shown in Fig.(\ref{fig:FR}). The time arrow in Fig.(\ref{fig:FR}) and throughout this paper is taken from left to right. Here are three more constraints in building Feynman diagrams with Fig.(\ref{fig:FR}): One, the energy conserves at each vertices. Two, take the integral of all the undermined energy appear in loop diagrams as ${\int}\frac{dE}{2\pi}$. And three, the energy of the incoming and outgoing lines is conserved. Whose derivation can also be found in Appendix A.

\section{Dressing effects}
        In classical field theory, say, electrodynamics, the dynamics of the field is given by Maxwell equation when the motion of source is specified. However, the classical electrodynamics fails give the dynamics of the source under the action of the field \cite{CED,Zhang}. This difficulty is inherited by the semiclassical theory where the field are introduced as quantum variables while the source are described by classical variables \cite{Itykson}. This implies a full quantum theory is needed to give a self-consistent description of the backaction effects. In quantum electrodynamics, the dynamics of electrons are also quantized. As a result of which, bare electrons travelling through the electromagnetic vacuum will get dressed with a cloud of virtual photons, which gives self-consistence modifications to the physical quantities of the electrons such as mass, charge and spin \cite{Itykson,Stephano,Peskin}. In this section, we will discuss the dressing effects of the photons, phonon and the coupling strength in our system, among which the dressing effect of phonon is the first backaction effect we encounter.

    \subsection{Frequency shift of harmonic oscillator}
\begin{figure}
\begin{center}
\includegraphics[width=.45\textwidth]{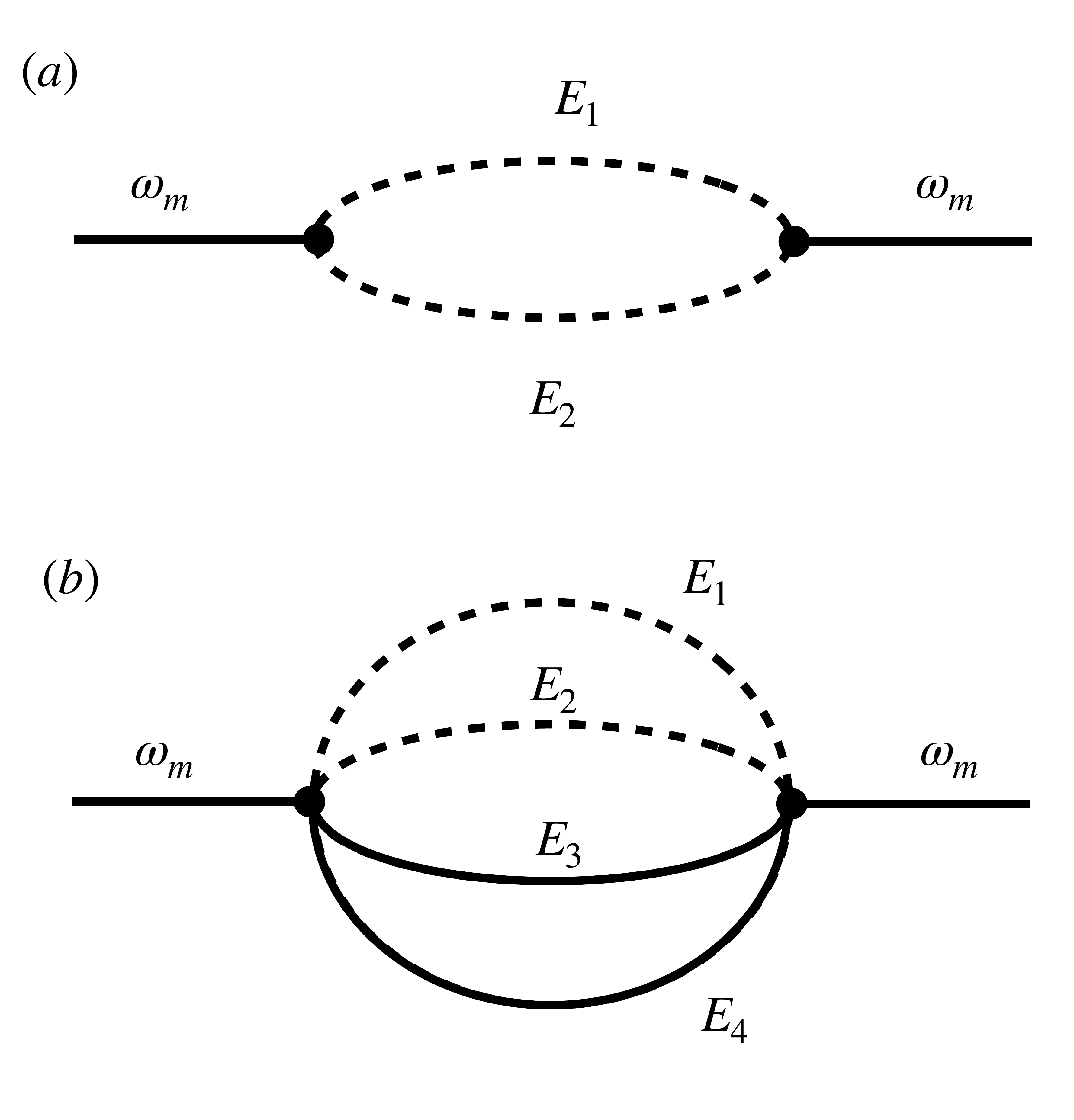}
\caption{One particle irreducible diagrams of phonon propagator. (a) Diagram of two virtual photons. (b) Diagram of four virtual photons.}
\label{fig:m}
\end{center}
\end{figure}

        In quantum electrodynamics, the loop diagrams of the fermion lines gives the electromagnetic mass. In this part, we will mimic such procedure. To the lowest order of $g$, there exists two one particle irreducible diagrams shown in Fig.(\ref{fig:m}). Under the chain approximation \cite{Peskin}, the frequency shift $\delta\omega_{m}$ is given as

        \begin{equation}\label{eq:m}
        \delta\omega_{m}=\Sigma_{2}^{(2)}(\omega_{m})+\Sigma_{2}^{(4)}(\omega_{m}),
        \end{equation}
        where $-i\Sigma_{2}^{(2)}(\omega_{m})$ and $-i\Sigma_{2}^{(4)}(\omega_{m})$ are the values of Fig.(\ref{fig:m}a) and Fig.(\ref{fig:m}b), respectively. The superscripts denote the order of diagrams while the subscripts denotes the number of intermediate particles.

        With the Feynman rules, the value of Fig.(\ref{fig:m}a) is
        \begin{equation}\label{eq:m2}
        \begin{split}
        -i\Sigma_{2}^{(2)}(\omega_{m})&=-\frac{g^{2}}{2}\int_{-\infty}^{+\infty}\frac{dE_{1}dE_{2}}{2\pi}\frac{i}{E_{1}-\omega_{c}+i\epsilon}\frac{i}{E_{2}-\omega_{c}+i\epsilon}\delta(E_{1}+E_{2}-\omega_{m})\\
        &=i\frac{g^{2}}{2}\frac{1}{2\omega_{c}-\omega_{m}}.
        \end{split}
        \end{equation}
        The integral in Eq.(\ref{eq:m2}) is convergent. And the value of Fig.(\ref{fig:m}b) is
        \begin{equation}\label{eq:m4}
        \begin{split}
        -i\Sigma_{4}^{(2)}(\omega_{m})&=-g^{2}\int_{-\infty}^{+\infty}\frac{dE_{1}dE_{2}dE_{3}dE_{4}}{(2\pi)^{3}}\frac{i}{E_{1}-\omega_{c}+i\epsilon}\frac{i}{E_{2}-\omega_{c}+i\epsilon}\frac{i}{E_{3}-\omega_{m}+i\epsilon}\\
        &\times\frac{i}{E_{4}-\omega_{b}+i\epsilon}\delta(E_{1}+E_{2}+E_{3}+E_{4}-\omega_{m})\\
        &=ig^{2}\frac{1}{2\omega_{c}+\omega_{m}}.
        \end{split}
        \end{equation}
        This integral is also convergent. When the coupling $g$ is sufficient small, the frequency shift of harmonic oscillator Eq.\eqref{eq:m} vanishes due to the $V_{DCE}$ being negligible. In the rotation wave approximation applied in \cite{EPL,Italian}, only diagram Fig.(\ref{fig:m}a) survives.

        Taking Eqs.(\ref{eq:m2}) and (\ref{eq:m4}) into Eq.(\ref{eq:m}) we obtain the frequency shift of harmonic oscillator
        \begin{equation}\label{eq:omegam}
        \delta\omega_{m}=-g^{2}(\frac{1}{2}\frac{1}{2\omega_{c}-\omega_{m}}+\frac{1}{2\omega_{c}+\omega_{m}}),
        \end{equation}
        which is identical what one can get from perturbation method to the state $\ket{0,1}$, where the first number in the ket denotes photon number while the second number denotes phonon number. From Fig.(\ref{fig:m}) we can see that the cloud of virtual particles, both photon and phonon shifts the phonon frequency. As a result, the observed frequency $\Omega_{m}=\omega_{m}+\delta\omega_{m}$ is no longer only determined by the strength of the spring and the mass of mirror, but also by the length of the cavity and the cavity field frequency.

        Besides the correction to the physical quantities, another consequence of the dressing effect is the renormalization of the field strength. In \cite{EPL,Italian}, the mean value of which is used to model the dynamical Casimir effect.

        The field equation of the phonon field obtained by Heisenberg motion equation is
        \begin{equation}
        \frac{db}{dt}=-i[\omega_{m}b+ga^{\dagger}a+\frac{g}{2}[a^{2}+(a^{\dagger})^{2}],
        \end{equation}
        the solution of which can be formally written as
        \begin{equation}\label{eq:bt}
        b(t)=b_{0}(t)+\int dt'G_{r}(t-t')f(t'),
        \end{equation}
        where $b_{0}(t)$ is the solution of the homogenous motion equation Eq.\eqref{eq:homo} and the $G_{r}(t-t')$ is the retarded Green function of the phonon field, the source term $f(t)$ is
        $$f=-i[ga^{\dagger}a+\frac{g}{2}[a^{2}+(a^{\dagger})^{2}].$$
        Eq.\eqref{eq:bt} gives us a picturized description of a physical phonon being composed of a bare phonon with a cloud of virtual particles. To keep in consistent with canonical quantization condition, the interacting field equation Eq.\eqref{eq:bt} must satisfy the boson statistics $[b,b^{\dagger}]=1$ which was imposed on the free field $b_{0}$ originally. As a result of which, the free field $b_{0}$ takes a percentage of $Z_{b}$ of the physical field $b$ and is no longer normalized. That is why the coefficient $Z_{b}$ called the field strength normalization factor. The field strength renormalization factor $Z_{b}$ can be computed via
        \begin{equation}\label{eq:Zb}
        \begin{split}
        Z_{b}^{-1}&=1-\frac{d\Sigma(\Omega_{m})}{dE}\\
        &=1+g^{2}[\frac{1}{2(2\omega_{c}-\omega_{m})^{2}}-\frac{1}{(2\omega_{c}+\omega_{m})^{2}}].
        \end{split}
        \end{equation}

        From Eqs.(\ref{eq:omegam}) and \eqref{eq:Zb} we can see $\Omega_{m}$ and $Z_{b}$ go divergent at resonance $\omega_{m}=2\omega_{c}$. However, this does not cause any inconsistency. The reason is, when tuned in resonance, the dynamical Casimir effect term $V_{DCE}$ will introduce process $\text{b}{\rightarrow}2\text{a}$, which is forbidden by energy conservation if there is detuning. This means the phonon is no longer stable in resonance. As a result of which, the energy of the phonon will no longer be determined but will experience a distribution width due to the time-energy uncertainty \cite{QM}. The detailed computation of the decay width and the life time of the phonon in resonance will be given in the next section.

    \subsection{Frequency shift of cavity field}
\begin{figure}
\begin{center}
\includegraphics[width=.45\textwidth]{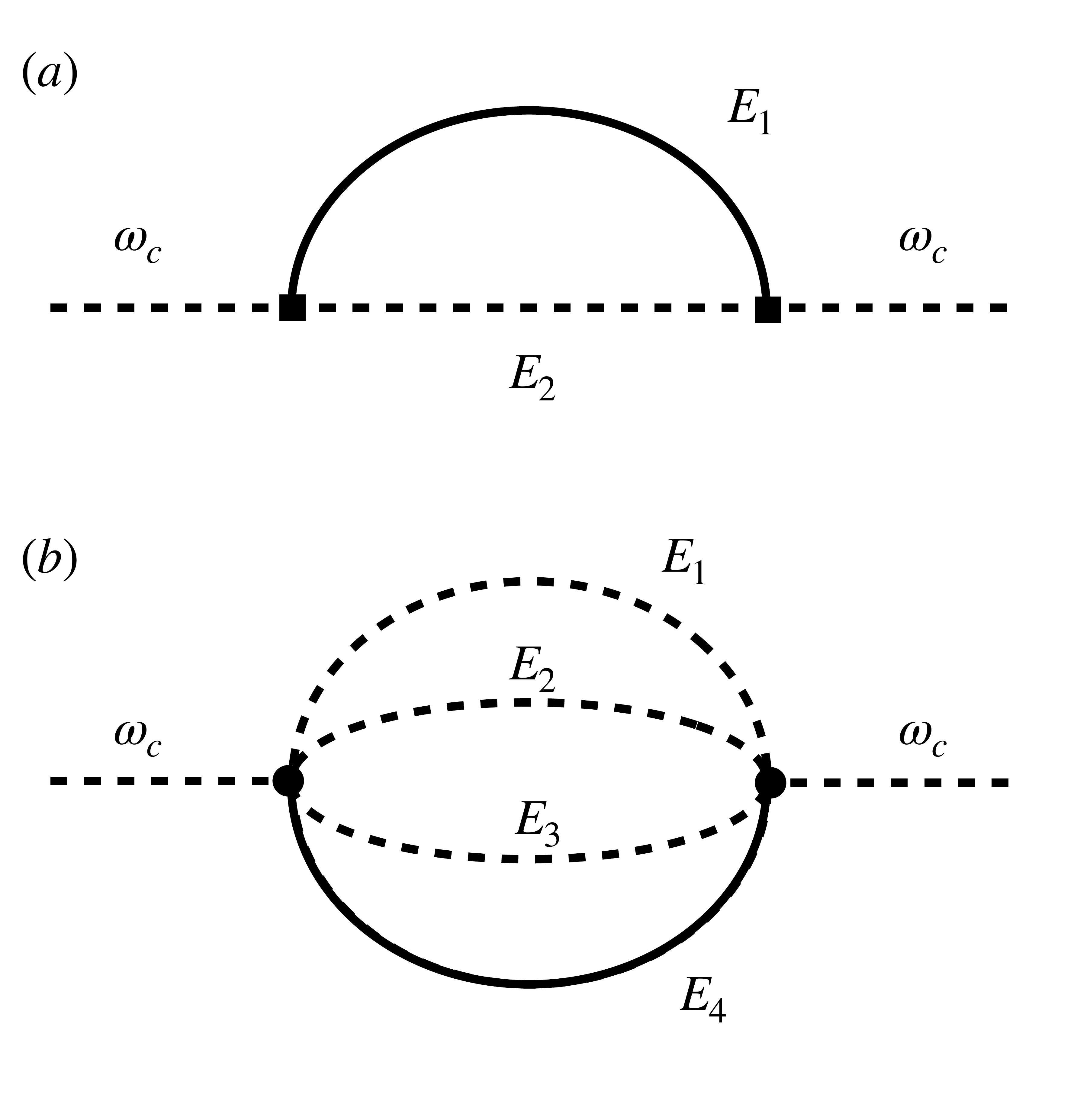}
\caption{One particle irreducible diagrams of photon propagator. (a) Diagram with two virtual particles.  (b) Diagram with four virtual particles.}
\label{fig:c}
\end{center}
\end{figure}

        To the lowest order of $g$, there are two one particle irreducible diagrams of photon lines demonstrated in Fig.(\ref{fig:c}), which gives the frequency shift as
        \begin{equation}\label{eq:c}
        \delta\omega_{c}=\Pi^{(2)}_{om}(\omega_{c})+\Pi^{(2)}_{DCE}(\omega_{c}),
        \end{equation}
        where $-i\Pi^{(2)}_{om}(\omega_{c})$ and $-i\Pi^{(2)}_{DCE}(\omega_{c})$ are the values of Fig.({\ref{fig:c}}a) and Fig.({\ref{fig:c}}b), respectively. The superscript denotes the order of the diagrams while the subscripts $om$ and $DCE$ denote the type of vertices. Here we note that Fig.(\ref{fig:c}a) survives when $V_{DCE}$ becomes negligible and the photon frequency shift Eq.\eqref{eq:c} vanishes in the rotation wave approximation \cite{EPL,Italian}.

        The values of the diagrams in Fig.(\ref{fig:c}(a)) and Fig.(\ref{fig:c}(b)) are
        \begin{equation}
        \begin{split}
        -i\Pi^{(2)}_{om}(\omega_{c})&=-g^{2}\int_{-\infty}^{\infty}\frac{dE_{1}dE_{2}}{2\pi}\frac{i}{E_{1}-\omega_{c}+i\epsilon}\frac{i}{E_{2}-\omega_{m}+i\epsilon}\delta(E_{1}+E_{2}-\omega_{c})\\
        &=ig^{2}\frac{1}{\omega_{m}},\\
        -i\Pi^{(2)}_{DCE}(\omega_{c})&=-\frac{3g^{2}}{2}\int_{-\infty}^{+\infty}\frac{dE_{1}dE_{2}dE_{3}dE_{4}}{(2\pi)^{3}}\frac{i}{E_{1}-\omega_{c}+i\epsilon}\frac{i}{E_{2}-\omega_{c}+i\epsilon}\frac{i}{E_{3}-\omega_{c}+i\epsilon}\frac{i}{E_{4}-\omega_{m}+i\epsilon}\\
        &\times\delta(E_{1}+E_{2}+E_{3}+E_{4}-\omega_{c})\\
        &=i\frac{3g^{2}}{2}\frac{1}{2\omega_{c}+\omega_{m}},
        \end{split}
        \end{equation}
        respectively. The integrals in the equations are convergent. Thus we obtain the frequency shift of photons
        \begin{equation}
        \delta\omega_{c}=-g^{2}(\frac{1}{\omega_{m}}+\frac{3}{2}\frac{1}{2\omega_{c}+\omega_{m}}).
        \end{equation}
        This result is identical to that from perturbation method to state $\ket{1,0}$. Like the picture we introduced in the dress effect of the phonon, a physical photon also contains a bare photon dressed in a cloud of virtual particles, whose field strength normalization factor is given by
        \begin{equation}
        \begin{split}
        Z_{a}^{-1}&=1-\frac{d\Pi(\Omega_{c})}{dE}\\
        &=1-\frac{3g^{2}}{(2\omega_{c}+\omega_{m})^{2}}.
        \end{split}
        \end{equation}

    \subsection{Vertex functions}
\begin{figure}
\begin{center}
\includegraphics[width=.45\textwidth]{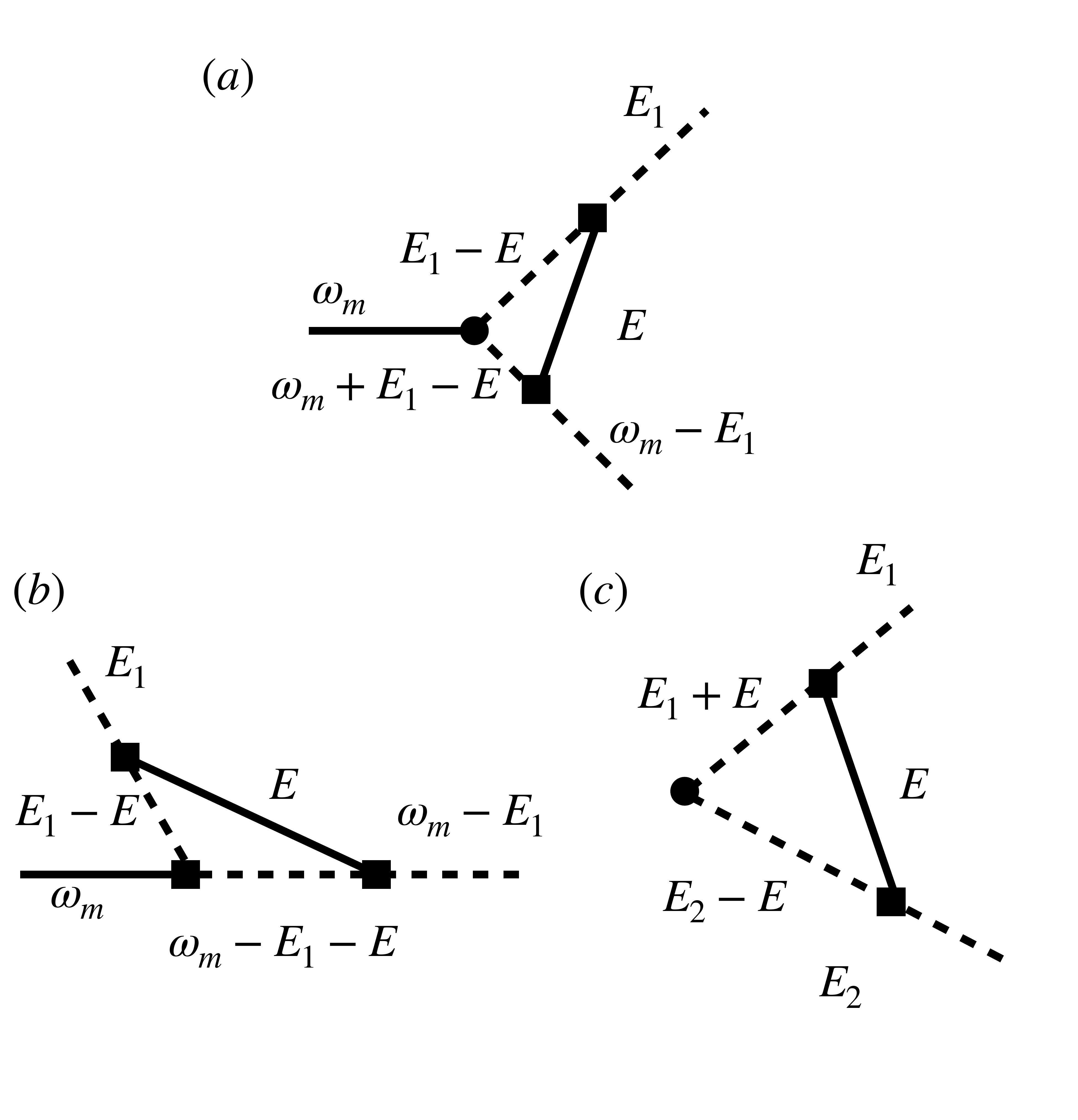}
\caption{Irreducible vertex loop diagrams. (a) Loop diagram of dynamical Casimir effect interaction vertex.  (b) Loop diagram of optomechanical interaction vertex. (c) Loop diagram of dynamical Casimir effect vertices with more than two virtual photon lines. Diagram (c) gives correction to Figs.(\ref{fig:m}(b)) and (\ref{fig:c}(b)).}
\label{fig:vertex}
\end{center}
\end{figure}

        The irreducible vertex loop diagrams to the lowest order of $g$ is given in Figs.(\ref{fig:vertex}a) and (\ref{fig:vertex}b), from which we can see that the optomechanical interaction plays central part in this effect.

        The corresponding vertex functions are
        \begin{equation}
        \begin{split}
        \Gamma_{DCE}(E_{1})&=\frac{ig^{3}}{4}\int\frac{dE}{2\pi}\frac{i}{E-\omega_{m}+i\varepsilon}\frac{i}{(E_{1}-E)-\omega_{c}+i\varepsilon}\frac{i}{(\omega_{m}+E-E_{1})-\omega_{c}+i\varepsilon}\\
        &=-\frac{ig^{3}}{4}\frac{1}{(E_{1}-\omega_{c}-\omega_{m})(\omega_{m}-2E_{1})},\\
        \Gamma_{opto}(E_{1})&=-\frac{ig^{3}}{8}\int\frac{dE}{2\pi}\frac{i}{E-\omega_{m}+i\varepsilon}\frac{i}{(E_{1}-E)-\omega_{c}+i\varepsilon}\frac{i}{(\omega_{m}+E-E_{1})-\omega_{c}+i\varepsilon}\\
        &=\frac{ig^{3}}{8}[\frac{1}{(E_{1}-\omega_{c}-\omega_{m})(2E_{1}-\omega_{m})}+\frac{1}{(\omega_{c}+E_{1})(\omega_{m}-2E_{1})}],
        \end{split}
        \end{equation}
        respectively, where the variable in the $\Gamma$ function denotes the energy of the outgoing lines in the diagrams with undetermined energy.  Thus the correction to the coupling strength are given by $\Gamma_{DCE}(E_{1})$ and $\Gamma_{opto}(E_{1})$, respectively.

        In diagrams with many virtual particles like Fig.(\ref{fig:m}b) and Fig.(\ref{fig:c}b) we will encounter the type of vertex graph in Fig.(\ref{fig:vertex}c)
        \begin{equation}
        \begin{split}
        \Gamma_{x}(E_{1},E_{2})&=\frac{ig^{3}}{4}\int\frac{dE}{2\pi}\frac{i}{E-\omega_{m}+i\varepsilon}\frac{i}{E_{1}+E-\omega_{c}+i\varepsilon}\frac{i}{E_{2}-E-\omega_{c}+i\varepsilon}\\
        &=\frac{ig^{3}}{4}\frac{1}{(E_{1}-\omega_{c}-\omega_{m})(E_{1}+E_{2}-2\omega_{c})},
        \end{split}
        \end{equation}
        where there are two outgoing lines with undetermined energy marked by $E_{1}$ and $E_{2}$.

    \subsection{Vacuum effect}
\begin{figure}
\begin{center}
\includegraphics[width=.35\textwidth]{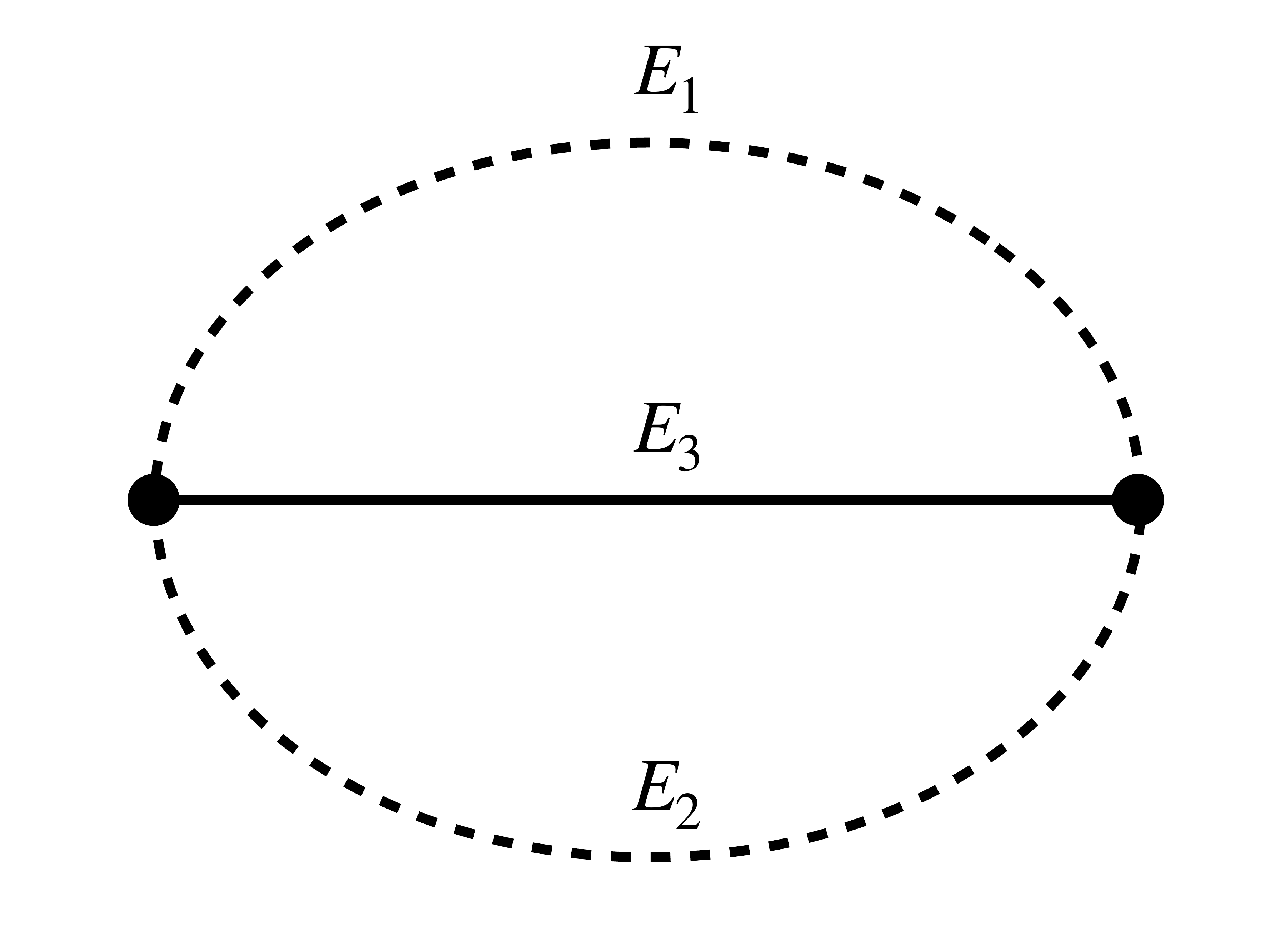}
\caption{The vacuum-vacuum diagram.}
\label{fig:vv}
\end{center}
\end{figure}

        In ground state, the quantum harmonic oscillator is not at still but exhibits zero-point motion, which squeezes the vacuum and may also generate radiation. Due to the energy conservation, the radiated particles cannot be real. In this part we will make use of the vacuum diagrams among which the one of lowest order of $g$ is shown in Fig.(\ref{fig:vv}). Note that this effect is purely induced by dynamical Casimir effect interaction.

        The value of Fig.(\ref{fig:vv}) is
        \begin{equation}
        \begin{split}
        -i\Lambda&=-\frac{g^{2}}{2}\int_{-\infty}^{+\infty}\frac{dE_{1}dE_{2}dE_{3}}{(2\pi)^{2}}\frac{i}{E_{1}-\omega_{c}+i\epsilon}\frac{i}{E_{2}-\omega_{c}+i\epsilon}\frac{i}{E_{3}-\omega_{m}+i\epsilon}\delta(E_{1}+E_{2}+E_{3})\\
        &=i\frac{g^{2}}{2}\frac{1}{2\omega_{c}+\omega_{m}},
        \end{split}
        \end{equation}
        which gives rise to the shift of the system's ground state energy by
        \begin{equation}\label{eq:v}
        {\delta}E_{g}=\Lambda.
        \end{equation}
        There is a sketchy way of understanding this, we pretend there is a propagator for the vacuum whose one particle irreducible loop diagram is Fig.(\ref{fig:vv}), then the shift of the vacuum energy can be obtained by chain approximation.

        Eq.\eqref{eq:v} is an exact match to that from perturbation theory to $\ket{0,0}$ and verifies the result in \cite{L111}. And the integral in Eq.\eqref{eq:v} is convergent. This can be interpreted as the zero-point movement of the mirror will generate "virtual radiation".

        When more than one mode of the cavity field is taken into consideration, the photons of the different modes can interact through the vacuum fluctuation of the phonon field. In this case, the interaction term becomes
        \begin{equation}\label{eq:Vij}
        V=\sum_{ij}g_{ij}(b+b^{\dagger})a_{i}^{\dagger}a_{j}+\frac{g_{ij}}{2}(b+b^{\dagger})(a_{i}a_{j}+a_{i}^{\dagger}a_{j}^{\dagger}),
        \end{equation}
        where $a_{i}$ and $a_{i}^{\dagger}$ denote the annihilation and creation operator of the cavity field mode $i$ and $g_{ij}=(\frac{1}{2})^{\frac{3}{2}}(-1)^{i+j}\frac{1}{L\sqrt{m}}\sqrt{\frac{\omega_{i}\omega_{j}}{\omega_{m}}}$ denotes the coupling strength when the photon lines of mode $i$ and $j$ attach to the vertex. In deriving Eq.\eqref{eq:Vij}, the normal ordering technique is applied to get rid of the infinite vacuum energy \cite{D91,L111}.

\begin{figure}
\begin{center}
\includegraphics[width=.45\textwidth]{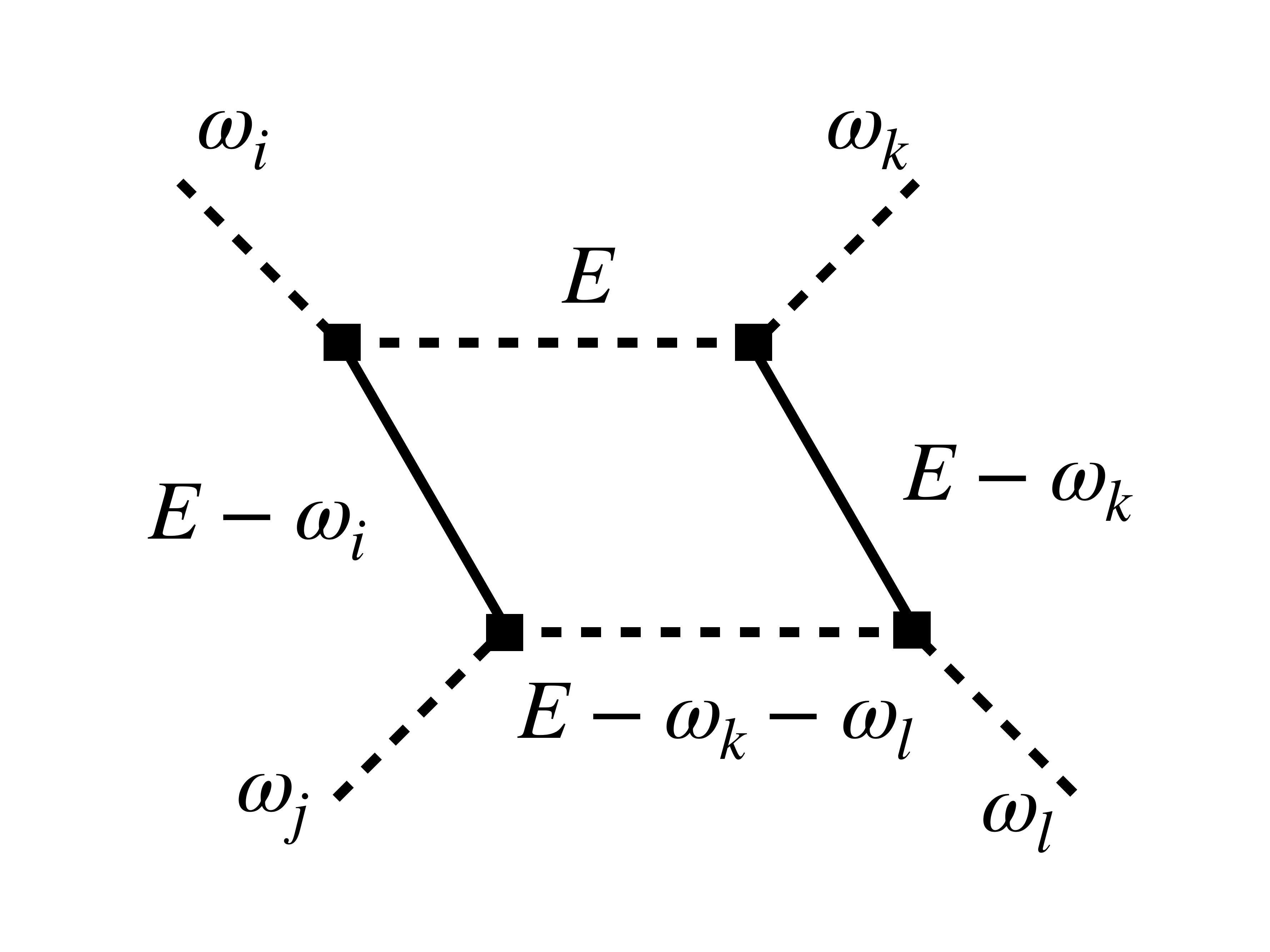}
\caption{The photon-photon interaction diagram, where the photon propagator can be of any mode.}
\label{fig:vv2}
\end{center}
\end{figure}

        To the lowest order of $g$, the effective photon-photon interaction diagram is shown in Fig.(\ref{fig:vv2}). We can see this effect is induced solely by optomechanical interaction. The value of Fig.(\ref{fig:vv2}) is thus given by

        \begin{equation}
        \begin{split}
        \sigma&=\sum_{ab}(-ig_{ab})^{4}\int\frac{dE}{2\pi}\frac{i}{E-\omega_{i}-\omega_{m}+i\varepsilon}\frac{i}{E-\omega_{k}-\omega_{m}+i\varepsilon}\frac{i}{E-\omega_{z}+i\varepsilon}\frac{i}{E-\omega_{k}-\omega_{c}-\omega_{b}+i\varepsilon}\\
        &=\sum_{ab}ig_{ab}^{4}[\frac{1}{(\omega_{i}-\omega_{k})(\omega_{i}+\omega_{m}-\omega_{a})(\omega_{i}+\omega_{m}-\omega_{k}-\omega_{l}-\omega_{b})}\\
        &+\frac{1}{(\omega_{k}-\omega_{i})(\omega_{k}+\omega_{m}-\omega_{a})(\omega_{m}-\omega_{l}-\omega_{b})}+\frac{1}{(\omega_{a}-\omega_{i}-\omega_{m})(\omega_{a}-\omega_{k}-\omega_{m})(\omega_{a}-\omega_{k}-\omega_{c}-\omega_{b})}\\
        &+\frac{1}{(\omega_{j}+\omega_{b}-\omega_{m})(\omega_{i}+\omega_{j}+\omega_{b}-\omega_{k}-\omega_{m})(\omega_{i}+\omega_{j}+\omega_{b}-\omega_{a})}].
        \end{split}
        \end{equation}
        Note here we ignored the Boson statistics of the outgoing photons and this is a process of order $g^{4}$.

\section{Dynamical Casimir effect and backaction}
        The dynamical Casimir effect is energy transferring from the mirror to the field in nature. In scattering theory language, this can be modeled by the processes $r\text{b}{\rightarrow}s\text{a}$ when parameters are properly tuned as $r\omega_{m}=s\omega_{a}$, where $r$ and $s$ are the number of the phonon and photon involved. Throughout this section we will be working in tree-level diagrams.

\begin{figure}
\begin{center}
\includegraphics[width=.42\textwidth]{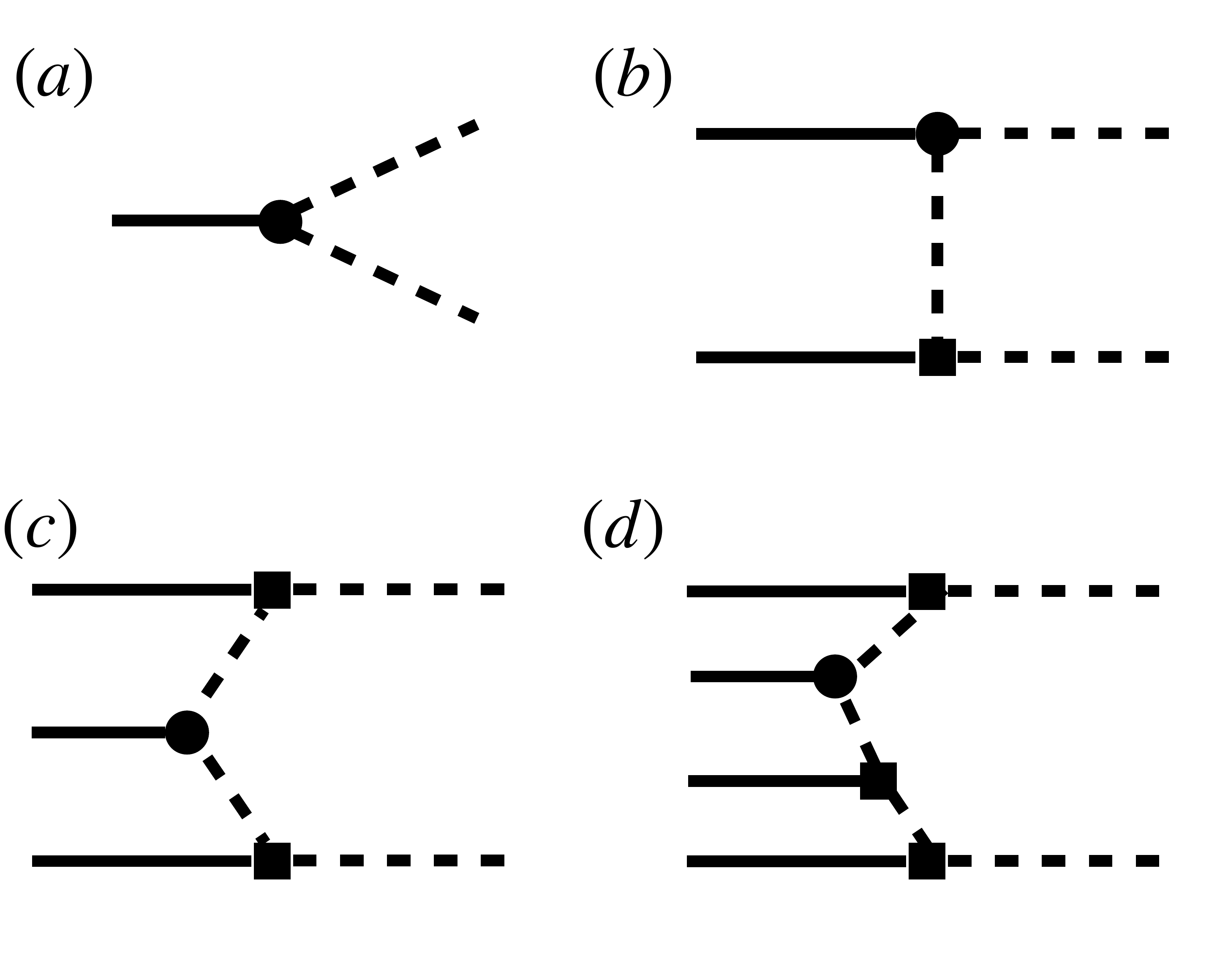}
\caption{Diagrams of the lowest order dynamical Casimir effect process: (a) Diagram of decay of a phonon, (b) Diagram of $2\text{b}{\rightarrow}2\text{a}$ process, (c) Diagram of $3\text{b}{\rightarrow}2\text{a}$ process, (d) Diagram of $4\text{b}{\rightarrow}2\text{a}$ process. Note that we dropped the markings for energies because they can be easily be deduced by the energy conservation rules at each vertices and of the whole process.}
\label{fig:dce}
\end{center}
\end{figure}

        In Fig.(\ref{fig:dce}) we presented some processes of dynamical Casimir effect to the lowest order of $g$. From which we can see that the there are always two phonon taking part in the dynamical Casimir effect, i.e. $r=2$. First we consider the process $\text{b}\rightarrow 2\text{a}$ shown in Fig.(\ref{fig:dce}a). The decay width of the phonon can be directly seen from Fig.(\ref{fig:dce}a) as $$\Gamma=g^{2}.$$
        Thus the life time of phonon is obtained
        \begin{equation}
        \tau=\frac{1}{\Gamma}=\frac{1}{g^{2}}.
        \end{equation}

        The diagrams of $k\text{b}{\rightarrow}2\text{a}$ with $k=2,3,4$ are shown in Figs.(\ref{fig:dce}b), (\ref{fig:dce}c) and (\ref{fig:dce}d), respectively. From which we can deduce that for processes $k\text{b}{\rightarrow}2\text{a}$ with arbitrary $k$, there will be $k-1$ optomechanical vertices and one dynamical Casimir effect vertices in the diagrams, thus the corresponding transition amplitude
        \begin{equation}\label{eq:amplitudes}
        A_{k{\rightarrow}2}\sim\frac{1}{2^{k-1}}g^{k}.
        \end{equation}

        Since the coupling constant $g$ is small, the transition amplitude Eq.\eqref{eq:amplitudes} decrease rapidly with growing $k$. This is qualitatively in agreement with the result obtained in the previous work \cite{PRX}.

        When photons are generated out of the vacuum, they take the energy away from the mirror. In the meantime, they exert radiation pressure on the mirror. This is the most commonly known backaction effect. The statistical nature of the backaction force has been studied persistently \cite{scattering,82,Neto94,Unruh14,Unruhmassive,1998,France} and the debate on whether the mirror will exhibit Brownian motion has also lasted for a very long time \cite{Unruh14,1998,France}. In the following part, we will use the Feynman diagram approach to explore the statistic nature of the backaction force of our system.

        The basic tool will be used is the correlation function of two operators $A$ and $B$
        \begin{equation}\label{eq:corr}
        \bra{\Omega}T[A(t_{1})B(t_{2})]\ket{\Omega}=\frac{\bra{0,0}T[U_{I}(+\infty,t_{1})A_{I}(t_{1})U_{I}(t_{1},t_{2})B_{I}(t_{2})U_{I}(t_{2},-\infty)]\ket{0,0}}{\bra{0,0}U_{I}(+\infty,-\infty)\ket{0,0}},
        \end{equation}
        where $T$ is the time-ordering symbol, $U_{I}(t)$ is the evolution operator in interaction picture and $A$ and $B$ are arbitrary time-dependent operators, $\ket{\Omega}$ is the ground state of the full Hamiltonian.
        
\begin{figure}
\begin{center}
\includegraphics[width=.48\textwidth]{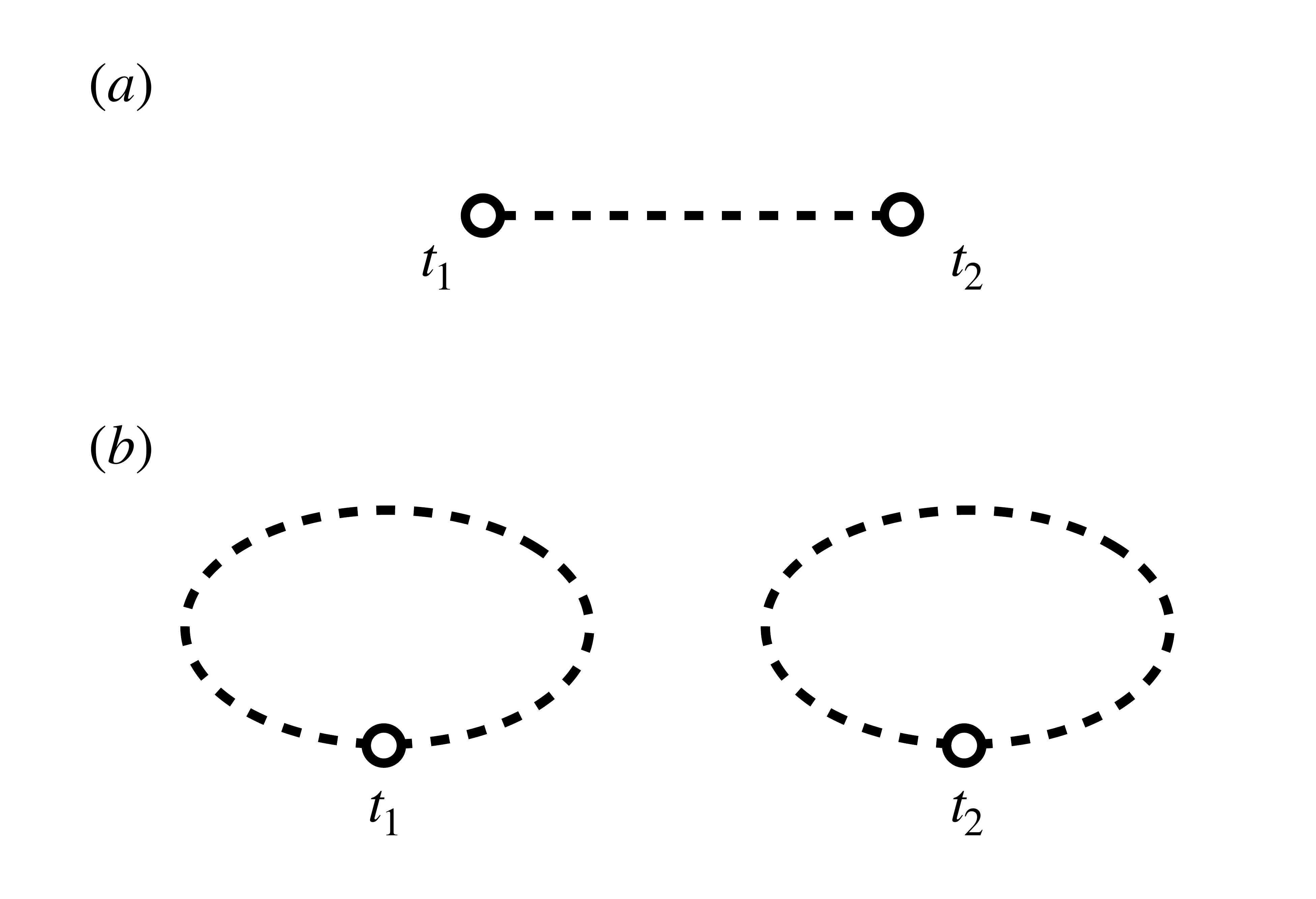}
\caption{Diagrams of the force-force correlation function to the lowest order: (a) connected diagram, (b) unconnected diagram. The empty circles at the end of the photon lines denotes they are not of interaction origin.}
\label{fig:ff}
\end{center}
\end{figure}

        Taking $A$ and $B$ as the backaction force operator $F$ in Eq.\eqref{eq:force} we obtain the its correlation function. Power expand the evolution operator gives the force-force correlation function
        \begin{equation}\label{eq:ff}
        \bra{\Omega}T[F(t_{1})F(t_{2})]\ket{\Omega}=\frac{\omega_{c}^{2}}{4L^{2}}e^{-2i\omega_{c}(t_{1}-t_{2})}
        \end{equation}
        to the lowest order of $g$, the corresponding diagrams are given in Fig.(\ref{fig:ff}),where $G_{0}(t)$ is the free Green function of Eq.\eqref{eq:H0}. Note that the unconnected diagram Fig.(\ref{fig:ff}b) is ruled out by the denominator in Eq.\eqref{eq:corr}. Eq.\eqref{eq:ff} demonstrates that the backaction force at different time have non-vanishing correlations and is of order $L^{-3}$, which is a small number. This means the Langevin-type equations serves as a approximation of modeling the motion of the mirror to the order of $L^{-3}$. The techniques used in deriving Eq.\eqref{eq:ff} can be found in Appendix A.

\section{Conclusion and Discussion}
        In this paper we have developed the Feynman diagram technique to study the fully quantized optomechanical system. We studied the dressing effects of phonon, photon and strength coupling, and demonstrated how the vacuum fluctuation modifies the physical quantities and physical particles of the system. The phenemenon of non-zero photon population will modify the ground state energy is verified and modeled as virtual radiation process. The dynamical Casimir effect is modeled elegantly by the scattering process converting the phonon pair into photons and the corresponding transition rates are derived. We also revealed the non-Gaussian nature of the possibility distribution of the backaction force and showed the Langivan-type equation serves as a good approximation of modeling the mirror's motion.

        The optimechanical system studied in this paper is conceptual a model of interaction between the mechanical degree of freedom and quantized field, thus implying many similarity to the black hole with Hawking radiation. Unfortunately, due to the lack of a manifestable quantum gravity theory, we can now modeling the black hole evaporation only on a semiclassical level \cite{HR2}. The techniques developed and new effects revealed in our system may be of help in the further investigation of the quantum effects in black hole evaporation, through which the road towards the theory of quantum gravity may be lighted \cite{QG}.

\subsection*{Acknowledgements}
        We thank W. H. Sang for comments on early drafts of this paper. Y. S. Cao thanks K. F. Lyu for discussion about the loop diagram computation. We thank R. Zhang for useful discussion.

\appendix
\section{Green function and Feynman rules}
        The motion equation of free field $a$ dictated by Eq.\eqref{eq:H0} is
        \begin{equation}\label{eq:homo}
        (\frac{d}{dt}+i\omega_{c})a=0,
        \end{equation}
        the corresponding Green function is
        \begin{equation}\label{eq:Ga}
        G_{a}(t-t')=\int\frac{dE}{2\pi}\frac{i}{E-\omega_{c}+i\varepsilon}e^{-iE(t-t')},
        \end{equation}
        which is in consistence with the one obtained from Huygens principle
        $$G_{a}(t-t')=\theta(t-t')\bra{0}a(t)a^{\dagger}(t')\ket{0}$$.
        In energy representation
        \begin{equation}
        G_{a}[E]=\frac{i}{E-\omega_{c}+i\varepsilon},
        \end{equation}
        where we used parenthesis and square brackets to distinguish the primitive and Fourier transformed functions.

        The free Green operator of the system satisfies
        \begin{equation}
        (\frac{d}{dt}-iH_{0})G_{0}(t-t')=\delta(t-t'),
        \end{equation}
        whose solution is obtained as \cite{QM}
        \begin{equation}
        G_{0}(t-t')=\theta(t-t')e^{-iH_{0}(t-t')},
        \end{equation}
        where $\theta(x)$ is the step function. This gives us the relation between Green operator and time free evolution operator $U_{0}(t)$ as 
        \begin{equation}\label{eq:Greenoperator}
        G(t-t')=\theta(t-t')U_{0}(t-t').
        \end{equation}

        The free Green function Eq.\eqref{eq:Ga} can thus be written as
        \begin{equation}
        G_{a}(t-t')=\theta(t-t')\bra{1}G_{0}(t-t')\ket{1}.
        \end{equation}
        For multi-particle propagation we have
        \begin{equation}\label{eq:multiparticle}
        G_{a}^{n}(t-t')=\theta(t-t')\bra{n}G_{0}(t-t')\ket{n}.
        \end{equation}

        The $S$-matrix
        \begin{equation}\label{eq:Smatrix}
        \begin{split}
        S&=U_{I}(+\infty,-\infty)\\
        &=\sum_{n=0}^{\infty}\frac{(-i)^{n}}{n!}T\int dt_{1}dt_{2}\cdots dt_{n-1}dt_{n}[V(t_{1})V(t_{2})\cdots V(t_{n-1})V(t_{n})],
        \end{split}
        \end{equation}
        where $T$ is time-ordering symbol and $U_{I}(t,t')$ is the time evolution operator in interaction picture.

        Express the arbitrary $S$-matrix element in Eq.(\ref{eq:Smatrix}) in Schrodinger picture, its general term formula reads
        \begin{equation}\label{eq:generalterm}
        \begin{split}
        &\frac{(-i)^{n}}{n!}\bra{\alpha}T\int dt_{1}dt_{2}\cdots dt_{n-1}dt_{n}[V(t_{1})V(t_{2})\cdots V(t_{n-1})V(t_{n})]\ket{\beta}\\
        &=\frac{(-i)^{n}}{n!}\bra{\alpha}\int dt_{1}dt_{2}\cdots dt_{n-1}dt_{n}[G_{0}(+\infty,t_{1})VG_{0}(t_{1},t_{2})\cdots G_{0}(t_{n-1},t_{n})VG_{0}(t_{n},-\infty)]\ket{\beta},
        \end{split}
        \end{equation}
        where two facts used in deriving the formula above. First, state $\ket{\alpha}$ obeys the following rule evolving in time
        $$\ket{\alpha,t}=U_{0}^{\dagger}(t,t')\ket{\alpha,t'},$$
        and so is for $\ket{\beta}$. Second, the explicit form the Green operator in Eq.(\ref{eq:Greenoperator}) allows us to eliminate the time-ordering symbol in the second row due to the step function in Green operator.

        For a general term in Eq.(\ref{eq:generalterm}), the Fourier transformation
        \begin{equation}
        \begin{split}
        &\int dt_{1}dt_{2}\cdots dt_{n-1}dt_{n}[G_{0}(+\infty,t_{1})VG_{0}(t_{1},t_{2})\cdots G_{0}(t_{n-1},t_{n})VG_{0}(t_{n},-\infty)]\\
        &=\frac{1}{2\pi}\int dE_{1}dE_{2}\cdots dE_{n-1}dE_{n}G_{0}[E_{1}]VG_{0}[E_{2}]\cdots G_{0}[E_{n-1}]VG_{0}[E_{n}]\delta(E_{1}-E_{2})\delta(E_{2}-E_{3})\cdots \delta(E_{n-1}-E_{n})
        \end{split}
        \end{equation}
        gives its formula in energy representation. The reason why the interaction $V$ does not contain energy variable $E$ can be understood by the whole system being closed, which means the energy is conserved in each interaction vertices.
        
        From Eq.(\ref{eq:multiparticle}), we have
        \begin{equation}
        \bra{n}G_{0}[E]\ket{n}=\int\frac{dE_{1}}{2\pi}\frac{dE_{2}}{2\pi}\cdots\frac{dE_{n-1}}{2\pi}\frac{dE_{n}}{2\pi}G_{a}[E_{1}]G_{a}[E_{2}]\cdots G_{a}[E_{n-1}]G_{a}[E_{n}]\delta(\sum_{i=1}^{n}E_{i}-E),
        \end{equation}
        whose diagrammatic representation is the Feynman rules introduced in Sec. 2.


\end{document}